%Paper: hep-th/9506140
%From: GASPERINI@to.infn.it
%Date: Wed, 21 Jun 1995 15:38:22 +0300 (MET-DST)

\magnification=1200
\hsize 15true cm \hoffset=0.5true cm
\vsize 23true cm
\baselineskip=10pt

%minore o circa uguale
\def\laq{\raise 0.4ex\hbox{$<$}\kern -0.8em\lower 0.62
ex\hbox{$\sim$}}
%maggiore o circa uguale
\def\gaq{\raise 0.4ex\hbox{$>$}\kern -0.7em\lower
0.62 ex\hbox{$\sim$}}

\font\small=cmr8 scaled \magstep0

\font\medio=cmr10 scaled \magstep2
\outer\def\beginsection#1\par{\medbreak\bigskip
      \message{#1}\leftline{\bf#1}\nobreak\medskip\vskip-
\parskip
      \noindent}

\def \pa {\partial}
\def \ra {\rightarrow}

\def \ti {\tilde}
\def \la {\lambda}

\def \Da {\Delta}
\def \b {\beta}
\def \a {\alpha}
\def \ap {\alpha^\prime}

\def \ga {\gamma}
\def \sg {\sigma}
\def \da {\delta}

\def \r {\rho}

\def \e {\eta}
\def \es {\eta_s}
\def \om {\omega}
\def \Om {\Omega}
\def \noi {\noindent}

\def\sqr#1#2{{\vcenter{\hrule height.#2pt\hbox{\vrule
width.#2pt
height#1pt \kern#1pt\vrule width.#2pt}\hrule height.#2pt}}}

\def\lsim{\mathrel{\rlap{\lower4pt\hbox{\hskip1pt$\sim$}}
    \raise1pt\hbox{$<$}}}         %less than or approx. symbol
\def\gsim{\mathrel{\rlap{\lower4pt\hbox{\hskip1pt$\sim$}}
    \raise1pt\hbox{$>$}}}         %greater than or approx. symbol

\nopagenumbers
\line{\hfil  DFTT-38/95}
\line{\hfil June 1995}
\line{\hfil hep-th/9506140}

\vskip 3 cm
\centerline {\medio AMPLIFICATION OF VACUUM FLUCTUATIONS}
\vskip 0.5 true cm
\centerline{{\medio IN STRING COSMOLOGY
BACKGROUNDS}\footnote{*}{\small Talk given at
the {\it ``{\bf 3rd Colloque Cosmologie}"} (Observatoire de
Paris, 7-9 June 1995). To appear in the {\bf Proceedings}
(World Scientific, Singapore), ed. by H. de Vega and N.
Sanchez.}}

\vskip 1.5true cm
\centerline{M.Gasperini}
\bigskip
\centerline{\it Dipartimento di Fisica Teorica, Universit\`a di
Torino,} \centerline{\it Via P.Giuria 1, 10125 Turin, Italy,}
\centerline{\it and INFN, Sezione di Torino, Turin, Italy}

\vskip 1.5 true cm
\centerline{\medio Abstract}
\bigskip
\noindent
Inflationary string cosmology backgrounds can amplify
perturbations in a more efficient way than conventional
inflationary backgrounds, because the perturbation amplitude
may grow - instead of being constant - outside the horizon. If
not gauged away, the growing mode can limit the range of
validity of a linearized description of perturbations. Even in
the restricted linear range, however, this enhanced
amplification may lead to phenomenological consequences
unexpected in the context of the standard inflationary
scenario. In particular, the production of a relic graviton
background strong enough to be detected in future by LIGO,
and/or the generation of a stochastic electromagnetic
background strong enough to seed the cosmic magnetic fields
and to be responsible for the observed large scale anisotropy.

\vfill\eject

\footline={\hss\rm\folio\hss}
\pageno=1

\centerline{\bf AMPLIFICATION OF VACUUM
FLUCTUATIONS}
\centerline{\bf IN STRING COSMOLOGY BACKGROUNDS}
\bigskip \centerline{M. GASPERINI}
\centerline{\small{\it Dipartimento di Fisica Teorica,
Universit\`a di Torino, }}
\centerline{\small{\it Via P.Giuria 1, 10125 Turin, Italy,}}
\centerline{\small{\it and INFN, Sezione di Torino, Turin, Italy}}
\bigskip
\centerline{ABSTRACT}
\midinsert
\narrower
\noi
Inflationary string cosmology backgrounds can amplify
perturbations in a more efficient way than conventional
inflationary backgrounds, because the perturbation amplitude
may grow - instead of being constant - outside the horizon. If
not gauged away, the growing mode can limit the range of
validity of a linearized description of perturbations. Even in
the restricted linear range, however, this enhanced
amplification may lead to phenomenological consequences
unexpected in the context of the standard inflationary
scenario. In particular, the production of a relic graviton
background strong enough to be detected in future by LIGO,
and/or the generation of a stochastic electromagnetic
background strong enough to seed the cosmic magnetic fields
and to be responsible for the observed large scale anisotropy.
\endinsert
\bigskip
\centerline{\bf Table of Contents}
\bigskip
\baselineskip=15pt
\item{1.} Introduction.

\item{2.} The ``growing mode" problem.

\item{3.} ``Thermal" graviton spectrum from dilaton-driven
inflation.

\item{4.} Two-parameter model of background evolution.

\item{5.} Parameterized graviton spectrum.

\item{6.} Parameterized electromagnetic spectrum.

6.1. Seed magnetic fields.

6.2. The CMB radiation and its anisotropy.

\item{7.} Conclusions.

\vskip 1 cm
\noi
{\bf 1. Introduction.}
\baselineskip=10pt
\bigskip
\noi
It is well known that the time-evolution of a cosmological
background can amplify quantum fluctuations and generate
primordial perturbation spectra [1]. Such a parametric
amplification can be described, in a second-quantization
language, as the production of pairs from the vacuum under
the action of an external ``pump" field. But it may also
visualized, from a kinematic point of view, as a ``stretching"
process in which the comoving amplitude stays
frozen (instead of decreasing adiabatically), while
perturbations propagate under some effective potential
barrier.

This is typically what happens in the standard inflationary
scenario. In the inflationary backgrounds obtained from
the string effective action, however, the perturbation
amplitude may even grow [2] (instead of being frozen) during
the stretching period. This leads to a more efficient
amplification of perturbations, but the amplitude could grow
too much, in such backgrounds, so as to prevent us from
applying the standard linearized formalism.

In view of this aspect, the aim of this paper is twofold. On one
hand I want to show that in some case this anomalous growth
can be gauged away, so that perturbations can consistently
linearized in an appropriate frame. I will discuss, in particular,
the growing mode of scalar metric perturbations in a
dilaton-driven background, which appears in the standard
longitudinal gauge and which seems to complicate the
computation of the spectrum [3]. On the other hand I want to
show that, even if the growth is physical, and we have to
restrict ourself to a reduced portion of parameter space in
order to apply a linearized approach, such enhanced
amplification is nevertheless rich of interesting
phenomenological consequences.

I will discuss in particular three points: {\it i)} the production
of a relic gravity wave background with a spectrum strongly
enhanced in the high frequency sector, and its possible
observation by large interferometric detectors [4] (such as
LIGO and VIRGO); {\it ii)} the amplification of electromagnetic
perturbations due to their direct coupling to the dilaton
background, and the generation of primordial ``seeds" for the
galactic and extragalactic magnetic field [5]; {\it iii)} the
generation of the large scale CMB anisotropy directly from the
vacuum fluctuations of the electromagnetic field [6].

Throughout this paper the evolution of perturbations will be
discussed in a type of background to which I shall refer to, for
short, as ``string cosmology background". At low energy such
background represents a solution [7,8,9] of the tree-level,
zeroth order in $\ap$, gravi-dilaton action
$$
-\int d^{d+1}x \sqrt{|g|} e^{-\phi}\left(R+\pa_\mu\phi\pa^\mu
\phi\right) \eqno(1.1)
$$
(possibly complemented by string matter sources). It describes
the accelerated evolution from the string perturbative
vacuum, with flat metric and vanishing dilaton coupling
($\phi=-\infty$), towards a phase driven by the kinetic energy
of the dilaton field ($H^2\sim \dot\phi^2$), with negligible
contribution from the dilaton potential. In this initial phase the
curvature scale $H^2$ and the dilaton coupling $e^\phi$ are
both growing, at a rate uniquely determined by
the action (1.1).

The background can be consistently described in terms of the
action (1.1), however, only up to the time $t=t_s$ when the
curvature reaches the string scale, namely when $H\simeq
H_s= (\ap)^{-1/2}\equiv \la_s^{-1}$. At that time all orders in
$\ap$ (i.e. all higher-derivative corrections) become important,
and the background enters a truly ``stringy" phase, whose
kinematic details cannot be predicted on the ground of the
previous simple action. The presence of this high-curvature
phase cannot be avoided, as it is required [10] to stop the
growth of the curvature, to freeze out the dilaton, and to
arrange a smooth transition (at $t=t_1$) to the standard
radiation-dominated evolution (where $\phi=$const).

In previous works (see for instance [11]) we assumed that the
time scales $t_s$ and $t_1$ (marking respectively the
beginning of the string and of the radiation era) were of the
same order, and we computed the perturbation spectrum in
the sudden approximation, by matching directly the radiation
era to the dilaton-driven phase. Here I will consider a more
general situation in which the duration of the string era
($t_1/t_s$) is left completely arbitrary, and I will discuss its
effects on the perturbation spectrum.

The evolution from the flat and cold initial state to the highly
curved (and strongly coupled) final regime was previously
called "pre-big-bang" [8], in order to stress the
complementarity of that phase with respect to the standard,
post-big-bang, decelerated scenario. During such a
pre-big-bang epoch the accelerated evolution of the
background can be invariantly characterized, from a kinematic
point of view, as a phase of shrinking event horizons [2,8,9].
If, in particular, we parameterize the pre-big-bang scale factor
(in cosmic time) as
$$
a(t)\sim (-t)^\b, \,\,\,\,\,\,\,\,\,\,\, -\infty < t < 0 \eqno(1.2)
$$
the existence condition for shrinking event horizons
$$
\int_t^0 {dt'\over a(t')} < \infty \eqno(1.3)
$$
imposes $\b<1$. So there are two classes of backgrounds in
which the event horizon is shrinking.

For $\b<0$ we have a metric describing a phase of
accelerated expansion and growing curvature,
$$
\dot a >0 , \,\,\,\,\,\,\, \ddot a >0 , \,\,\,\,\,\,\, \dot H>0
\eqno(1.4)
$$
of the type of pole-inflation [12], also called super-inflation
($H=\dot a/a$, and a dot denotes differentiation with respect to
the cosmic time $t$). For $0<\b<1$ we have instead a metric
describing accelerated contraction and growing curvature
scale,
$$
\dot a <0 , \,\,\,\,\,\,\, \ddot a <0 , \,\,\,\,\,\,\, \dot H<0
\eqno(1.5)
$$
The first type of metric provides a representation of the
pre-big-bang scenario in the String (or Brans-Dicke) frame, in
which test strings move along geodesic surfaces. The second in
the Einstein frame, in which the gravi-dilaton action is
diagonalized in the standard canonical form.

In both types of backgrounds the computation of the metric
perturbation spectrum may become problematic, but the best
frame to illustrate the difficulties is probably the Einstein
frame, where the metric is contracting. It should be recalled,
in this context, that the tensor perturbation spectrum for
contracting backgrounds was first given by Starobinski [13],
but the possible occurrence of problems, due to a
growing solution of the perturbation equations, was pointed
out only much later [14], in the context of dynamical
dimensional reduction. The problem, however, was left
unsolved.
\vskip 2 cm
\noi
{\bf 2. The ``growing mode" problem.}
\bigskip
\noi
Consider the evolution of tensor metric perturbations, $\da
g_{\mu\nu}=a^2h_{\mu\nu}$, in a $(3+1)$-dimensional
conformally flat background, parameterized in conformal time
($\eta=\int dt/a$) by the scale factor
$$
a(\eta)\sim (-\eta)^\a, \,\,\,\,\,\,\,\,\,\,\, -\infty < \eta < 0
\eqno(2.1)
 $$
The Fourier components $u_k$ of the correct variable obeying
canonical commutation relations
($u_{\mu\nu}=aM_ph_{\mu\nu}$, $M_p$ is the Planck mass)
satisfy, for each of the two physical (transverse traceless)
polarizations, the well known perturbation equation [1]
$$
u_k''+(k^2-{a''\over a})u_k=0\eqno(2.2)
$$
(a prime denotes differentiation with respect to $\eta$). In a
string cosmology background the horizon is shrinking, so that
all comoving length scales $k^{-1}$ are ``pushed out"of the
horizon. For a mode $k$ whose wavelength is larger than the
horizon size (i.e. $|k\eta|<<1$), we have then the asymptotic
solution
$$
h_k= {u_k\over a M_p}=A_k+B_k|\eta|^{1-2\a} , ~~~~~~~~
\,\,\,\,\, \eta \ra0_- \eqno(2.3)
$$
where $A_k$ and $B_k$ are integration constants.

The asymptotic behavior of the perturbation is thus
determined by $\a$. If $\a <1/2$ the perturbation tends to stay
constant outside the horizon, and the typical amplitude
$|\da_h|$ at the scale $k^{-1}$, for modes normalized to an
initial vacuum fluctuation spectrum,
$$
\lim_{\eta \ra -\infty} u_k \sim {1\over \sqrt k} e^{-ik\eta}
\eqno(2.4)
$$
can be given as usual [15] in terms of the Hubble factor at
horizon crossing ($k\eta \sim 1$)
$$
|\da_h|=k^{3/2}|h_k|\simeq \left( H\over
M_p\right)_{HC}\eqno(2.5)
$$
In this case the amplitude is always smaller than one provided
the curvature is smaller than Planckian (this case includes, in
particular, $\a<0$, namely all backgrounds describing
accelerated inflationary expansion, according to eq. (2.1)).

If, on the contrary, $\a>1/2$, the second term is the dominant
one in the solution (2.4), the perturbation amplitude tends to
grow outside the horizon,
$$
|\da_h|=k^{3/2}|h_k|\simeq \left (H\over
M_p\right)_{HC}|k\eta|^{1-2\a} , ~~~~~~~~
\,\,\,\,\,\, \eta \ra 0_-
\eqno(2.6)
$$
and may become larger than one, thus breaking the validity of
the perturbative approach. Otherwise stated: the energy
density (in critical units) stored in the mode $k$,
$\Om(k)=d(\r/\r_c)/d\ln k$, may become larger than one in
contrast with the hypothesis of negligible back-reaction of
perturbations on the initial metric.

One might think that this problem - due to the dominance of
the second term in eq. (2.3) - appears in the Einstein frame
because of the contraction, but disappears in the String frame
where the metric is expanding. Unfortunately this is not true
because, in the String frame, the different metric background
is compensated by a different perturbation equation, in such a
way that the perturbation spectrum remains exactly the same
[2,9].

This important property of perturbations can be easily
illustrated by taking, as a simple example, an isotropic solution
of the $(d+1)$-dimensional gravi-dilaton equations [2],
obtained from the action (1.1) complemented by a perfect gas
of long, stretched strings as sources (with equation of state
$p=-\r/d$).

In the Einstein frame the solution describes a contracting
background for $\eta \ra 0_-$,
$$
a=\left(-\eta\right)^{2(d+1)/(d-1)(3+d)} , \,\,\,\,\,\,
\phi = -{4d\over 3+d}\sqrt{2\over d-1} \ln (-\eta) \eqno (2.7)
$$
and the tensor perturbation equation
$$
h_k'' +(d-1){a'\over a}h_k' +k^2h_k=0 \eqno(2.8)
$$
has an asymptotic solution (for $|k\eta|<<1$) which grows,
according to eq. (2.3), as
$$
\lim_{\eta \ra 0_-} h_k \sim |\eta |^{(1-d)/(d+3)} \eqno (2.9)
$$
In the String frame the metric is expanding,
$$
\ti a=\left(-\eta\right)^{-2/(3+d)} , \,\,\,\,\,\,\,\,\,
\ti \phi = -{4d\over 3+d} \ln (-\eta) \eqno (2.10)
$$
but the perturbation is also coupled to the time-variation of
the dilaton background [16],
$$
h_k'' +\left[(d-1){\ti a'\over \ti a}- \ti \phi'\right]h_k'
+k^2h_k=0 \eqno(2.11)
$$
As a consequence, the explicit form of the perturbation
equation is exactly the same as before,
$$
h_k'' +{2(d+1)\over d+3} {h_k'\over \eta} +k^2h_k=0 \eqno(2.12)
$$
so that the solution is still growing, asymptotically, with the
same power as in eq. (2.9).

It may be noted that in the String frame the growth of
perturbations outside the horizon is due to the joint
contribution of the metric and of the dilaton background to the
"pump" field responsible for the parametric amplification
process [17]. Such an effect is thus to be expected in generic
scalar-tensor backgrounds, as noted also in [18]. The particular
example chosen above is not much relevant, however, for a
realistic scenario in which the phase of pre-big-bang inflation
is long enough to solve the standard cosmological problems. In
that case, in fact, all scales which are inside our present
horizon crossed the horizon (for the first time) during the
dilaton-driven phase or during the final string phase, in any
case when the contribution of matter sources was negligible
[9,11,19].

We shall thus consider, as a more significant (from a
phenomenological point of view) background, the vacuum,
dilaton-driven solution of the action (1.1), which in the Einstein
frame (and in $d=3$) can be explicitly written as
$$
a=(-\eta)^{1/2} , \,\,\,\,\,\,\,\,\,\,\,\, \phi=-\sqrt 3 \ln (-\eta) ,
\,\,\,\,\,\,\,\,\,\,\,\,  -\infty < \eta <0 \eqno(2.13)
$$
In such a background one finds that the growth of tensor
perturbations is simply logarithmic [3],
$$
|\da_h(\eta)|\simeq \left |H\over
M_p\right|_{HC}\ln|k\eta| \simeq {H_s\over
M_p}|k\eta_s|^{3/2}\ln|k\eta| ,~~~ \,\,
|k\eta_s|<1 , \,\, |\eta|>|\eta_s|
\eqno(2.14)
$$
so that it can be easily kept under control, provided the
curvature scale $H_s\sim (a_s\eta_s)^{-1}$ at the end of the
dilaton phase is bounded.

The problem, however, is with scalar perturbations, described
in the longitudinal gauge by the variable $\psi$ such that [15]
$$
(g_{\mu\nu}+\da g_{\mu\nu})dx^\mu dx^\nu=
a^2(1+2\psi)d\eta^2 -a^2(1-2\psi) (dx_i)^2\eqno(2.15)
$$
The canonical variable $v$ associated to $\psi$ is defined (for
each mode $k$) by [15]
$$
\psi_k= -{\phi'\over 4k^2M_p} \left(v_k\over a\right)'
\eqno(2.16)
$$
and satisfies a perturbation equation
$$
v_k''+(k^2-{a''\over a})v_k=0\eqno(2.17)
$$
which is identical to eq. (2.2) for the tensor canonical
variable, with asymptotic solution
$$
{v_k\over a}\simeq {1\over \sqrt k} {\ln |k\eta|\over a_{HC}} ,
\,\,\,\,\,\,\,\,\,\,\,~~~~~~~~~~~~~~ |k\eta|<<1 \eqno(2.18)
$$
Because of the different relation between canonical variable
and metric perturbation, however, it turns out that the
amplitude of longitudinal perturbations, normalized to an initial
vacuum fluctuation spectrum,
$$
\lim_{\eta \ra -\infty} v_k \sim {1\over \sqrt k} e^{-ik\eta}
\eqno(2.19)
$$
grows, asymptotically, like $\eta^{-2}$. We have in fact, from
(2.16),
$$
|\da_\psi(\eta)|= k^{3/2}|\psi_k|\simeq \left |H\over
M_p\right||k\eta| ^{-1/2} \simeq
 \left |H\over
M_p\right|_{HC}|k\eta| ^{-2}\simeq
$$
$$
\simeq \left (H_s\over
M_p\right){|k\eta_s|^{3/2}\over |k\eta|^2} \sim {1\over
\eta^2} , \,\,\,\,\,\,~~~~~~ \eta \ra 0_-
\eqno(2.20)
$$

This growth, as we have seen, cannot be eliminated by passing
to the String frame. Neither can be eliminated in a background
with a higher number of dimensions. In fact, in $d>3$, the
isotropic solution (2.13) is generalized as [9]
$$
a=(-\eta)^{1/(d-1)} , \,\,\,\,\,\,\,\,\,\,\,\, \phi=-\sqrt
{2d(d-1)} \ln a ,
 \,\,\,\,\,\,\,\,\,\,\,\,  -\infty < \eta <0 \eqno(2.21)
$$
and the scalar perturbation equation in the longitudinal gauge
[9]
$$
\psi_k'' +3(d-1){a'\over a}\psi_k' +k^2\psi_k=0 \eqno(2.22)
$$
has the generalized asymptotic solution
$$
\psi_k= A_k +B_k\eta a^{-3(d-1)} \eqno(2.23)
$$
By inserting the new metric (2.21) one thus finds the same
growing time-behavior, $\psi_k\sim \eta^{-2}$, exactly as
before. The same growth of $\psi_k$ is also found in
anisotropic, higher-dimensional, dilaton-dominated
backgrounds [3].

Because of the growing mode there is always (at any given
time $\eta$) a low frequency band for which $|\da_\psi
(\eta)|>1$. In $d=3$, in particular, such band is defined [from
eq.(2.20)] as $k<\eta^{-1}(H/M_p)^2$. For such modes the
linear approximation breaks down in the longitudinal gauge,
and a full non-linear treatment would seem to be required in
order to compute the spectrum. In spite of this conclusion, a
linear description of scalar perturbations may remain possible
provided we choose a different gauge, more appropriate to
linearization than the longitudinal one.

A first signal that a perturbative expansion around a
homogeneous background can be consistently truncated at the
linear level, comes from an application of the ``fluid flow"
approach [20,21] to the perturbations of a scalar-tensor
background. In this approach, the evolution of density and
curvature inhomogeneities is described in terms of two
covariant scalar variables, $\Da$ and $C$, which are gauge
invariant to all orders [22]. They are defined in terms of the
momentum density of the scalar field, $\nabla \phi$, of the
spatial curvature, $^{(3)}R$, and of their derivatives. By
expanding around our homogeneous dilaton-driven background
(2.13) one finds [3] for such variables the asymptotic solution
($|k\eta|<<1$), in the linear approximation,
$$
\Da_k = const , \,\,\,\,\,\,\,~~~~~~~~~~
c_k= const +A_k \ln |k\eta| \eqno(2.24)
$$
showing that they tend to stay constant outside the horizon,
with at most a logarithmic variation (like in the tensor case),
which is not dangerous.

As a consequence, the amplitude of density and curvature
fluctuations can be consistently computed in the linear
approximation (for all modes) in terms of $\Da$ and $C$, and
their spectral distribution (normalized to an initial vacuum
spectrum) turns out to be exactly the same as the tensor
distribution (2.14), which is bounded.

What is important, moreover, is the fact that such a spectral
distribution could also be obtained directly from the
asymptotic solution of the scalar perturbation equations in the
longitudinal gauge [3],
$$
\psi_k = c_1 \ln |k\eta| + {c_2\over \eta^2} \eqno (2.25)
$$
simply by neglecting the growing mode contribution (i.e.
setting $c_2=0$). This may suggests that such growing mode
has no direct physical meaning, and that it should be possible
to get rid of it through an appropriate coordinate choice.

A good candidate to do the job is what we have called [3]
off-diagonal gauge,
$$
(g_{\mu\nu}+\da g_{\mu\nu})dx^\mu dx^\nu=
a^2\left[(1+2\varphi)d\eta^2 -(dx_i)^2-2\pa_iB dx^i d\eta\right]
\eqno(2.26)
$$
which represents a complete choice of coordinates, with no
residual degrees of freedom, just like the longitudinal gauge.
In this gauge there are two variables for scalar perturbations,
$\varphi$ and $B$, and their asymptotic solution in the linear
approximation is [3]
$$
\varphi_k= c_1 \ln |k\eta| \sim \psi_k , \,\,\,\,\,\,~~~~
B_k ={c_2\over \eta} \sim \eta \psi_k ,  \,\,\,\,\,\,~~~~
(\pa B)_k \sim |k\eta| \psi_k \eqno(2.27)
$$
($c_1$ and $c_2$ are integration constants). The growing mode
is thus completely gauged away for homogeneous
perturbations (for which $\pa_i B=0$). It is still present for
non-homogeneous perturbations in the off-diagonal part of
the metric, but it is  ``gauged down" by the factor $k\eta$
which is very small, asymptotically.

Fortunately this is enough for the validity of the linear
approximation, as the amplitude of the off-diagonal
perturbation, in this gauge,
$$
|\da_B| \simeq |k\eta||\da _\psi|
 \simeq \left (H_s\over
M_p\right)|k\eta_s|^{1/2}\left|\eta_s\over \eta\right|
\eqno(2.28)
$$
stays smaller than one for all modes $k<|\eta_s|^{-1}$, and for
the whole duration of the dilaton-driven phase, $|\e|>|\es|$.
We have explicitly checked that quadratic corrections are
smaller than the linear terms in the perturbation equations,
but a full second order computation requires a further
coordinate transformation [3]. The higher order problem is very
interesting in itself, but a complete discussion of the problem is
outside the scope of this paper. Having established that the
vacuum fluctuations of the metric background, amplified by a
phase of dilaton-driven evolution, can be consistently described
(even in the scalar case) as small corrections of the
homogeneous background solution, let me discuss instead some
phenomenological consequence of such amplification. Scalar
perturbations and dilaton production were discussed in
[9,11,23]. Here I will concentrate, first of all, on graviton
production.

\vskip 2 cm
\noi
{\bf 3. ``Thermal" graviton spectrum from dilaton-driven
inflation.}
\bigskip
\noi
Consider the amplification of tensor metric perturbation in a
generic string cosmology background, of the type of that
described in Sect. 1. Their present spectral energy distribution,
$\Om (\om, t_0)$, can be computed in terms of the Bogoliubov
coefficient determining their amplification (see Set. 5 below),
or simply by following the evolution of the typical amplitude
$|\da_h|$ from the time of horizon crossing down to the
present time $t_0$. For modes crossing the horizon in the
inflationary dilaton-driven phase (i.e. for $t<t_s$), and
reentering the horizon in the decelerated radiation era
($t>t_1$), one easily finds, from eq. (2.14)
$$
\Om (\om, t)\equiv {\om\over \r_c}{d\r \over d\om} \simeq
|\da_h|^2\simeq A \Om_\ga \left(H_s\over M_p\right)^2
\left(\om\over \om_s\right)^3 \ln^2 \left(\om\over
\om_s\right) ,~~~~  \om <\om_s \eqno(3.1)
$$

Here $\om=k/a$ is the red-shifted proper frequency for the mode
$k$ at time $t$, $\r_c=M_p^2 H^2 $ is the critical energy
density, $\Om_\ga = (H_1/H)^2(a_1/a)^4 = \r_\ga /\r_c$ is the
radiation energy density in critical units, and $\om_s=
H_sa_s/a$ is the maximal amplified frequency during the
dilaton-driven phase. Finally, $A$ is a possible amplification
factor due to the subsequent string phase ($t_s<t<t_1$), in
case that the perturbation amplitude
grows outside the horizon (instead
of being constant) during such phase. This additional
amplification does not modify however the slope of the
spectrum, as we are considering modes that crossed the
horizon before the beginning of the string phase.

An important property of the spectrum (3.1) is the universality
of the slope $\om^3$ with respect to the total number $d$ of
spatial dimensions, and their possible anisotropy. Actually, the
spectrum is also duality-invariant [4], in the sense that it is the
same for all backgrounds, including those with torsion,
obtained via $O(d,d)$ transformations [24] from the vacuum
dilaton-driven background.

The spectrum (3.1) has also the same
slope (modulo
logarithmic corrections) as the low frequency part of
a thermal black body spectrum, which can be written (in critical
units) as
$$
\Om _T(\om, t)= {\om^4\over \r_c}{1\over e^{\om/ T} -1}
 \simeq B \Om_\ga
\left(H_s\over M_p\right)^2
\left(\om\over \om_s\right)^3 {T\over \om_s} ,  \,\,~~~~~
\om <T\eqno(3.2)
$$
Here $B=(H_s/H_1)^2(a_s/a_1)^4$ is a constant factor which
depends on the time-gap between the beginning of the string
phase and the beginning of the string era. We can thus
parameterize the graviton spectrum (3.1) in terms of an
effective temperature
$$
T_s=(A/B) \om_s \eqno (3.3)
$$
which depends on the initial curvature scale $H_s$, and on the
subsequent kinematic of the high-curvature string phase.

For a negligible duration of the string phase, $t_s\sim t_1$, we
have in particular $H_s \sim H_1 \sim M_p$, and the spectrum
(3.1) is peaked around a maximal amplified frequency $\om_s
\sim H_1 a_1/a \sim 10^{11}$Hz, while it is exponentially
decreasing at higher frequencies (where the parametric
amplification is not effective). Moreover, $T_s \sim \om_s
\sim 1 ^oK$, so that this spectrum, produced by a geometry
transition, is remarkably similar to that of the observed cosmic
black body radiation [25,26] (see also Sect. 6.2 below).

The problem, however, is that we don't know the duration and
the kinematics of the high curvature string phase. As a
consequence, we know the slope ($\om ^3$) of this ``dilatonic"
branch of the spectrum, but we don't know the position, in the
($\Om,\om$) plane, of the peak frequency $\om_s$. This
uncertainty is, however, interesting, because the effects of
the string phase could shift the spectrum (3.1) to a low enough
frequency band, so as to overlap with the possible future
sensitivity of
large interferometric detectors such as LIGO [27] and VIRGO
[28]. I will discuss this possibility in terms of a two-parameter
model of background evolution, presented in the following
Section.

\vskip 2 cm
\noi
{\bf 4. Two-parameter model of background evolution.}
\bigskip
\noi
Consider the scenario described in Sect. 1 (see also [9,11]),
in which the initial (flat and cold) vacuum state, possibly
perturbed by the injection of an arbitrarily small (but finite)
density of bulk string matter, starts an accelerated evolution
towards a phase of growing curvature and dilaton coupling,
where the matter contribution becomes eventually negligible
with respect to the gravi-dilaton kinetic energy. Such a phase
is initially described by the low energy dilaton-dominated
solution,
$$
a=|\e|^{1/2} , \,\,\,\,\,\,\,\,\,\,
\phi = -\sqrt 3 \ln |\e| , \,\,\,\,\,\,\,\,\,\,
-\infty < \e < \es \eqno(4.1)
$$
up to the time $\es$, when the curvature reaches the string
scale $H_s=\la_s^{-1}$, at a value of the string coupling $g_s=
\exp (\phi_s/2)$. Provided the value of $\phi_s$ is sufficiently
negative (i.e. provided the coupling $g_s$ is sufficiently
small to be still in the perturbative regime), such a value is
also completely arbitrary, since there is no perturbative
potential to break invariance under shifts of $\phi$.

For $\e >\es$ the background enters a high curvature string
phase of arbitrary (but unknown) duration, in which all
higher-derivative (higher-order in $\ap =\la_s^2$)
contributions to the effective action become important.
During such phase the dilaton keeps growing towards the
strong coupling regime, up to the time $\e=\e_1$ (at a
curvature scale $H_1$), when a non-trivial dilaton potential
freezes the coupling to its present constant value $g_1=
\exp(\phi_1/2)$. We shall assume, throughout this paper, that
the time scale $\e_1$ marks the end of the string era as well
as the (nearly simultaneous) beginning of the standard,
radiation-dominated evolution, where $a\sim \e$ and $\phi =$
const (see however the last comment at the end of Sect. 6.2 for a
possible alternative).

During the string phase the curvature is expected to stay
controlled by the string scale, so that
$$
|H|\simeq g M_p = {e^{\phi/2} \over \la_p} = {1\over \la_s} ,
\,\,\,\,\,\,\,~~~~~~ \es < \e < \e_1 \eqno (4.2)
$$
where $\la_p$ is the Planck length. As a consequence, the
curvature is increasing in the Einstein frame (where $\la_p$ is
constant), while it keeps constant in the string frame, where
$\la_s$ is constant and the Planck length grows like $g$ from
zero (at the initial vacuum) to its present value $\la_p\simeq
10^{-19}(GeV)^{-1}$. In both cases the final scale $H_1\simeq
g_1 M_p$ is fixed, and has to be of Planckian order to match
the present value of the ratio $\la_p/\la_s$. One can estimate
[29]
$$
g_1 \simeq {H_1\over M_p} \simeq {\la_p \over\la_s} \simeq
0.3 - 0.03 \eqno(4.3)
$$

In analogy with the dilaton-driven solution (4.1), let us now
parameterize, in the Einstein frame, the background kinematic
during the string phase with a monotonic metric and dilaton
evolution,
$$
a=|\e|^{\a} , \,\,\,\,\,\,\,\,\,\,
\phi = -2\b \ln |\e| , \,\,\,\,\,\,\,\,\,\,
\e_s< \e < \e_1 \eqno(4.4)
$$
representing a sort of ``average" time-behavior. Note that
the two parameters $\a$ and $\b$ cannot be independent
since, according to eq. (4.2),
$$
\left |H_s\over H_1\right| \simeq {g_s\over g_1} \simeq
\left |\e_1\over \es\right|^{1+\a}\simeq
\left |\e_1\over \es\right|^{\b} \eqno(4.5)
$$
from which
$$
1+\a \simeq \b \simeq -{\log (g_s/g_1)\over \log |\es/\e_1|}
\eqno(4.6)
$$
(note also that the condition $1+\a=\b$ cannot be satisfied by
the vacuum
solutions of the lowest order string effective action [9], in
agreement with the fact that all orders in $\ap$ are full
operative in the high curvature string phase [10]).

The background evolution, for this class of models, is thus
completely determined in terms of two parameters only, the
duration (in conformal time) of the string phase, $|\es/\e_1|$,
and the shift of the dilaton coupling (or of the curvature scale
in Planck units) during the string phase, $g_s/g_1=(H_s/M_p)/
(H_1/M_p)$. I will use, for convenience, the decimal logarithm
of these parameters,
$$
\eqalign {x&=\log_{10} |\es/\e_1|=\log_{10} z_s \cr
y&=\log_{10} (g_s/g_1)=\log_{10} {(H_s/M_p)\over (H_1/M_p)}
\cr }
\eqno(4.7)
$$
Here $z_s=|\es/\e_1|\simeq a_1/a_s$ defines the total
red-shift associated to the string phase in the String frame,
where the curvature is constant and the metric undergoes a
phase of de Sitter-like expansion. It should be noted, finally,
that the parameters (4.7) are completely frame-independent,
as conformal time and dilaton field are exactly the same in the
String and Einstein frame.

\vskip 2 cm
\noi
{\bf 5. Parameterized graviton spectrum.}
\bigskip
\noi
Consider the background discussed in the previous Section,
characterized by the dilaton-driven evolution (4.1) for
$\e<\es$, by the string evolution (4.4) for $\es<\e<\e_1$, and
by the standard radiation-dominated evolution for $\e>\e_1$.
In these three regions, eq. (2.2) for the canonical variable
$u_k$ has the general exact solution
$$
\eqalign{u_k&=|\eta|^{1/2}
H_{0}^{(2)}(|k\eta|), ~~~~~~~~~~~~~~~~~~~~~~~~~~~\,\,\,\,\,\,\,\,\,\,\,\,
{}~~~~~~~~~~~~~\eta<\eta_{s}\cr
u_{k} &= |\eta|^{1/2}\left[ A_{+}(k)
H_{\nu}^{(2)}(|k\eta|) + A_{-}(k)
H_{\nu}^{(1)} (|k\eta|)\right] ,~~~~~~~~~
\eta_{s}<\eta<\eta_{1} \cr
u_{k}&= {1\over \sqrt k}\left[c_{+}(k)e^{-ik\eta}
+c_{-}(k)e^{ik\eta}\right] ,~~~~~~~~~~~~~~~~~~~~~~~~~~~\eta>\eta_{1} \cr}
\eqno(5.1)
$$
where $\nu=|\a-1/2|$, and
$H^{(1,2)}$
are the first and second kind Hankel functions. We have
normalized the solution to an initial vacuum fluctuation
spectrum, containing only positive frequency modes at $\e
=-\infty$
$$
\lim_{\e \ra -\infty} u_k= {e^{-ik\e}\over \sqrt k} \eqno(5.2)
$$
The asymptotic solution for $\e \ra +\infty$ is however a linear
superposition of positive and negative frequency modes,
determined by the so-called Bogoliubov coefficients
$c_\pm(k)$ which parameterize, in a second quantization
approach, the unitary transformation connecting $|in\rangle $
and $|out \rangle$ states. So, even starting from an initial
vacuum state, it is possible to find a non-vanishing expectation
number of produced particles (in this case gravitons) in the
final state, given (for each mode $k$) by $\langle n_k \rangle=
|c_-(k)|^2$.

We shall compute $c_\pm$ by matching the solutions (5.1) and
their first derivatives at $\es$ and $\e_1$. We observe, first of
all, that a consistent growth of the curvature and of the coupling
during the string phase (in the Einstein frame) can only be
realized by choosing $|\e_1|<|\es|$, i.e. $\b=1+\a>0$ [see
eq.(4.5)]. This  corresponds to an inflationary string phase,
characterized in the Einstein frame by accelerated expansion
($\dot a>0$, $\ddot a >0$, $\dot H >0$) for $-1<\a<0$, and
accelerated contraction
($\dot a<0$, $\ddot a <0$, $\dot H <0$) for $\a>0$,.
It follows, in particular, that modes which ``hit"
the effective potential barrier $V(\e)=a''/a$ of eq.(2.2)
(otherwise stated: which cross the horizon) during the
dilaton-driven phase, i. e. modes with $|k\e_s|<1$, stay under
the barrier also during the string phase, since $|k\e_1|< |k\e_s|
<1$. In such case the maximal amplified proper frequency
mode
$$
\om_1 ={k_1\over a}\simeq {1\over a\e_1}
\simeq {H_1a_1\over a}\simeq \left(H_1\over M_p\right)^{1/2}
10^{11} Hz
=\sqrt {g_1} 10^{11} Hz \eqno(5.3)
$$
is related to the highest mode crossing the horizon in the
dilaton phase, $\om_s=H_s a_s/a$, by
$$
\om_s=\om_1 \left|\e_1\over \es\right| <\om_1 \eqno(5.4)
$$
For an approximate estimate of $c_-$ we may thus consider two
cases.

If $\om_s<\om < \om_1$, i.e. if we consider modes crossing the
horizon in the string phase, we can estimate $c_-(\om)$ by
using the large argument limit of the Hankel functions when
matching the solutions at $\e=\es$, using however the small
argument limit when matching at $\e=\e_1$. In this case the
parametric amplification is induced by the second background
transition only, as $A_+\simeq 1$ and $A_-\simeq 0$, and we
get
$$
|c_-(\om)|\simeq \left(\om\over
\om_1\right)^{-\nu-1/2} , \,\,\,\,\,\,\,\,\,\,~~~~~
\om_s<\om<\om_1
\eqno(5.5)
$$
(modulo numerical coefficients of order of unity). If, on the
contrary, $\om<\om_s$, i.e. we consider modes crossing the
horizon in the dilaton phase, we can use the small argument
limit of the Hankel functions at both the matching epochs $\es$
and $\e_1$. This gives $A_\pm=b_\pm |k\es|^{-\nu}\ln |k\es|$
($b_\pm$ are numbers of order one), and
$$
|c_-(\om)|\simeq \left |\es\over \e_1\right|^\nu
\left(\om\over
\om_1\right)^{1/2}\ln \left(\om_s\over\om\right) , \,\,\,\,\,
{}~~~~~\om<\om_s
\eqno(5.6)
$$

We can now compute, in terms of $\langle n \rangle=
|c_-|^2$, the spectral energy distribution $\Om(\om,t)$ (in
critical units) of the produced gravitons, defined in such a way
that the total graviton energy density $\r_g$ is obtained as
$\r_g=\r_c\int \Om(\om)d\om/\om$. We have then
$$
\eqalign{\Om(\om,t)&\simeq {\om^4\over
M_p^2H^2}|c_-(\om)|^2 \simeq \cr
&\simeq \left(H_1\over M_p\right)^2
\left(H_1\over H\right)^2\left(a_1\over a\right)^4
\left(\om\over \om_1\right)^{3-2\nu} , ~~~~~~
\,\,\,\,\,\,~~~ \om_s<\om<\om_1 \cr
&\simeq \left(H_1\over M_p\right)^2
\left(H_1\over H\right)^2\left(a_1\over a\right)^4
\left(\om\over \om_1\right)^{3} \left|\es\over
\e_1\right|^{2\nu}\ln^2\left(\om_s\over \om \right) ,~~~
\,\, \om< \om_s \cr}\eqno(5.7)
$$
According to eqs. (4.6) and (4.7), moreover, $2\nu=|2\a -1|=
|3+2y/x|$ and $|\es/\e_1|=10^x$. The tensor perturbation
spectrum (5.7) is thus completely fixed in terms of our two
free parameters, $x, y$, of the (known) fraction of critical
energy density stored in radiation at time $t$, $\Om_\ga (t)=
(H_1/H)^2(a_1/a)^4$, and of the (in principle known) present
value of the ratio $g_1=\la_p/\la_s$, as
$$
\Om(\om,t)=g_1^2\Om_\ga (t)\left(\om\over
\om_1\right)^{3-|{2y\over x}+3|} , ~~~~\,\,\,\,\,\, 10^{-x}<{\om
\over \om_1}<1  \eqno(5.8)
$$
$$
\Om(\om,t)=g_1^2\Om_\ga (t)\left(\om\over
\om_1\right)^{3}10^{|{2y}+3x|} , ~~~~\,\,\,\, \,\,{\om
\over \om_1}<10^{-x} \eqno(5.9)
$$

The first branch, with unknown slope, is due to modes crossing
the horizon in the string phase, the second to modes crossing
the horizon in the dilaton phase. Note that I have omitted, for
simplicity, the logarithmic term in eq. (5.9), because it is not
much relevant for the order of magnitude estimate that I want
to discuss here. Note also that the same spectrum can also be
obtained with a different approach, working in the String
frame (see [4]).

Let us impose, on such spectrum, the condition of falling within
the possible future sensitivity range of large interferometric
detectors, namely [30]
$$
\Om(\om_I) \gaq 10^{-10} , \,\,\,\,\,\,\,\,\,\,\,
\om_I =10^2 Hz \eqno(5.10)
$$
which implies
$$
|y+{3\over 2}x|> {3\over 2}x -{(3+\log_{10}g_1)x\over 9+
\log_{10}g_1} , \,\,\,\,\,\, x> 9+{1\over2}\log_{10}g_1
$$
$$
|2y+3x|>21-{1\over 2}\log_{10}g_1 , \,\,\,\,\,\,\,\,\,\,
x<9+{1\over 2}\log_{10}g_1 \eqno(5.11)
$$
These conditions define an allowed region in our
parameter space $(x,y)$, which has to be further restricted
however by the upper bound obtained from pulsar-timing
measurements [31], namely
$$
\Om(\om_P) \laq 10^{-6} , \,\,\,\,\,\,\,\,\,\,\,
\om_P =10^{-8} Hz \eqno(5.12)
$$
which implies
$$
|y+{3\over 2}x|< {3\over 2}x -{(1+\log_{10}g_1)x\over
19+{1\over 2} \log_{10}g_1} , \,\,\,\,\,\, x>
19+{1\over2}\log_{10}g_1
$$
$$
|2y+3x|<55-{1\over 2}\log_{10}g_1 , \,\,\,\,\,\,\,\,\,\,
x<19+{1\over 2}\log_{10}g_1 \eqno(5.13)
$$

We have to take into account, in addition, the asymptotic
behavior of tensor perturbations outside the horizon. During
the dilaton phase the growth is only logarithmic, but during
the string phase the growth is faster (power-like) for
backgrounds with $\a>1/2$. Since the above spectrum has been
obtained in the linear approximation, expanding around a
homogeneous background, we must impose for consistency
that the perturbation amplitude stays always smaller than
one, so that perturbations have a negligible back-reaction on
the metric. This implies $\Om<1$ at all $\om$ and $t$. This
bound, together with the slightly more stringent bound
$\Om<0.1$ required by standard nucleosynthesis [32], can be
automatically satisfied - in view of the $g_1^2$ factor in eqs.
(5.8), (5.9) - by requiring a growing perturbation spectrum,
namely
$$
y<0 , \,\,\,\,\,\,\,\,\,\,\,\,\,\,\,\,~~~~~~~~~~~
 y>-3x \eqno(5.14)
$$

The conditions (5.11), (5.13) and (5.14) determine the allowed
region of our parameter space compatible with the production
of cosmic gravitons in the interferometric sensitivity range (5.10)
(denoted by LIGO, for short). Such a region is plotted in {\bf
Fig.1}, by taking $g_1=1$ as a reference value. It is bounded
below by the condition of nearly homogeneous
background (5.14), and above by the same condition plus the
pulsar bound (5.13). The upper part of the allowed region
corresponds to a class of backgrounds in which the tensor
perturbation amplitude stays constant, outside the horizon,
during the string phase ($\a<1/2$). The lower part corresponds
instead to backgrounds in the the amplitude grows, outside the
horizon, during the string phase ($\a>1/2$).

We note, finally, that the area within the full bold lines refers
to modes crossing the horizon in the dilaton phase; the area
within the thin lines  refers
to modes crossing the horizon in the string phase, where the
reliability of our predictions is weaker, as we used
field-theoretic methods in a string-theoretic regime. Even
neglecting all spectra referring to the string phase, however,
we obtain a final allowed region which is non-vanishing,
though certainly not too large.

\vfill\eject

The main message of this figure and of the spectrum
(5.7) (irrespective of the particular value of the spectral
index) is that graviton production, in string cosmology, is in
general strongly enhanced in the high frequency sector
(kHz-GHz). Such a frequency band, in our context, could be in
fact all but the ``desert" of relic gravitational radiation that
one may expect on the ground of the standard inflationary
scenario. In particular, a sensitivity of $\Om \sim
10^{-4} - 10^{-5}$ in the KHz region (which does not seem out of reach
in coincidence experiments between bars and interferometers [33])
could be already
enough to detect a signal, so that a null result (in that
band, at that level of sensitivity) would already provide a
significant constraint on the parameters of the string
cosmology background. This should encourage the study and
the development of gravitational detectors (such as, for
instance, microwave cavities [34]) with large sensitivity in the
high frequency sector.

In the following Section I will compare the allowed region of
{\bf Fig. 1}, relative to graviton production (and their possible
detection), to the allowed region relative to the amplification
of electromagnetic perturbations (and to the production of
primordial magnetic fields).

\vskip 2 cm
\noi
{\bf 6. Parameterized electromagnetic spectrum}
\bigskip
\noi
In string cosmology, the electromagnetic field $F_{\mu\nu}$ is
directly coupled to the dilaton background. To lowest order,
such coupling is represented by the string effective action as
$$
\int d^{d+1}x\sqrt{|g|} e^{-\phi}F_{\mu\nu}F^{\mu\nu}
\eqno(6.1)
$$
The electromagnetic field is also coupled to the metric
background $g_{\mu\nu}$, of course, but in $d=3$ the
metric coupling is conformally invariant, so that no parametric
amplification of electromagnetic fluctuations is possible in a
conformally flat background, like that of a typical inflationary
model. One can try to break conformal invariance at the
classical or quantum level - there are indeed various attempts
in this sense [35,36] - but it turns out that it is very difficult, in
general, to obtain a significant electromagnetic amplification
from the metric coupling in a natural way, and in a realistic
inflationary scenario.

In our context, on the contrary, the vacuum fluctuations of the
electromagnetic field can be directly amplified by the time
evolution of the dilaton background [5, 37]. Consider in fact the
correct canonical variable $\psi^\mu$ representing
electromagnetic perturbations [according to eq. (6.1)] in a
$d=3$, conformally flat background, i.e. $\psi^\mu=A^\mu
e^{-\phi/2}$, where $F_{\mu\nu}= \pa_\mu A_\nu -
\pa_\nu A_\mu$. The Fourier modes $\psi^\mu_k$ satisfy, for
each polarization component, the equation
$$
\psi_k''+\left[k^2-V(\e)\right]\psi_k=0 , \,\,\,\,\,\,\,\,
{}~~~~~~V(\e)= e^{\phi/2}(e^{-\phi/2})'' \eqno(6.2)
$$
obtained from the action (6.1) with the gauge condition
$\pa_\nu (e^{-\phi}\pa^\mu A^\nu)=0$. Such equation is very
similar to the tensor perturbation equation (2.2), with the only
difference that the Einstein scale factor  is replaced by the
inverse of the string coupling, $g^{-1}=e^{-\phi/2}$.

Consider now the string cosmology background of Sect. 4, in
which the dilaton-driven phase (4.1) and the string phase (4.4)
are followed by the radiation-driven expansion. For such
background, the effective potential (6.2) is given explicitly by
$$
\eqalign{
V&={1\over {4\eta^2}} (3 -
\sqrt{12}) ,~~~~~~~~~~~~~~~\,\,\,\,\eta<\eta_{s}  \cr
V&={\beta(\beta-1)\over{\eta^2}} ,
{}~~~~~~~~~~~~~~~~\eta_{s}<\eta<\eta_{1} \cr
V&=0 ,~~~~~~~~~~~~~~~~~~~~~~~~~~~~~~~~~\eta>\eta_{1}\cr}
\eqno(6.3)
$$
The exact solution of eq. (6.2), normalized to an initial
vacuum fluctuation spectrum ($\psi_k \ra e^{-ik\e}/\sqrt k$ for
$\e \ra -\infty$), is thus
$$
\eqalign{\psi_k&=|\eta|^{1/2}
H_{\sg}^{(2)}(|k\eta|) ,~~~~~~~~~~~~~~~~~~~~\,\,\,\,\,\,\,\,\,\,\,\,
{}~~~~~~~~~~~~~~~~\eta<\eta_{s}\cr
\psi_{k} &= |\eta|^{1/2}\left[ B_{+}(k)
H_{\mu}^{(2)}(|k\eta|) + B_{-}(k)
H_{\mu}^{(1)} (|k\eta|)\right] ,~~~~~
\eta_{s}<\eta<\eta_{1} \cr
\psi_{k}&= {1\over \sqrt k}\left[c_{+}(k)e^{-ik\eta}
+c_{-}(k)e^{ik\eta}\right] ,~~~~~~~~~~~~~~~~~~~~~~~\eta>\eta_{1} \cr}
\eqno(6.4)
$$
where $\sg= (\sqrt 3 -1)/2$, and $\mu=|\b-1/2|$.

For this model of background evolution the effective potential
$V(\eta)$ grows in the dilaton phase, keeps growing in the
string phase where it reaches a maximum $\sim \e_1^{-2}$
around the transition scale $\e_1$, and then goes rapidly to
zero in the subsequent radiation phase, where
$\phi=\phi_1=$const. The maximum amplified frequency is of
the same order as before,
$\om_1=H_1a_1/a=|\es/\e_1|\om_s >\om_s$, where
$\om_s=H_sa_s/a$ is the last mode hitting the barrier (or
crossing the horizon) in the dilaton phase. For modes with
$\om>\om_s$ the amplification is thus due to the second
background transition only: we can evaluate $|c_-|$ by using
the large argument limit of the Hankel functions when
matching the solutions at $\es$ (which gives $B_+\simeq 1$,
$B_-\simeq 0$), using however the small argument limit when
matching at $\e_1$, which gives
$$
|c_-(\om)|\simeq \left(\om\over \om_1\right)^{-\mu-1/2} ,
\,\,\,\,\,\,\,\, \om_s<\om<\om_1 \eqno (6.5)
$$
Modes with $\om<\om_s$, which exit the horizon in the dilaton
phase, stay outside the horizon also in the string phase, so
that we can use the small argument limit at both the matching
epochs: this gives $B_\pm =b_\pm |k\es|^{-\sg -\mu}$
($b_\pm$ are numbers of order of unity) and
$$
|c_-(\om)|\simeq \left(\om \over \om_s \right)^{-\sg}
\left(\om\over \om_1\right)^{-1/2}\left|\es\over
\e_1\right|^{\mu} ,  \,\,\,\,\,\, \om<\om_s \eqno (6.6)
$$

We are interested, in particular, in the ratio
$$
r(\om)={\om\over \r_{\ga}}{d\r\over d\om}\simeq
{\om^4\over \r_{\ga}}|c_-(\om)|^2 \eqno(6.7)
$$
measuring the fraction of electromagnetic energy density
stored in the mode $\om$, relative to the total radiation
energy $\r_\ga$. By using the parameterization of Sect. 4 we
have $2\mu=|2\b-1|=|1+2y/x|$ and $|\es
/\e_1|=\om_1/\om_s=10^x$, so that the electromagnetic
perturbation spectrum is again determined by two parameters
only, the duration of the string phase $|\es/\e_1|$, and the
initial value of the string coupling, $g_s=g_1 10^y$. We find
$$
r(\om)=g_1^2\left(\om\over
\om_1\right)^{3-|{2y\over x}+1|} , ~~~~~
\,\,\,\,\,\, 10^{-x}<{\om
\over \om_1}<1  \eqno(6.8)
$$
for modes crossing the horizon in the string phase, and
$$
r(\om)=g_1^2\left(\om\over
\om_1\right)^{4-\sqrt 3}10^{x(1-\sqrt 3) +|{2y}+x|} ,
\,\,\,\,~~~~~
\,\,{\om \over \om_1}<10^{-x} \eqno(6.9)
$$
for modes crossing the horizon in the dilaton phase. The same
spectrum has been obtained, with a different approach, also in
the String frame [5,6]. Note that, in this paper, the definition of
$g_1$ has been rescaled with respect to [6], by absorbing into $g_1$ the
$4\pi$ factor.
\bigskip
\noi
{\bf 6.1. Seed magnetic fields}
\bigskip
\noi
The above spectrum of amplified electromagnetic vacuum
fluctuations has been obtained in the linear approximation,
expanding around a homogeneous background. We have thus to
impose on the spectrum the consistency condition of negligible
back-reaction, $r(\om)<1$ at all $\om$.  For $g_1<1$ this
condition requires a growing perturbation spectrum, and
imposes a rather stringent bound on parameter space,
$$
y<x , \,\,\,\,\,\,\,\,\,\,\,\,\,\,\,\,\, y>-2x \eqno(6.10)
$$
(note that a growing spectrum also automatically satisfies the
nucleosynthesis bound $r<0.1$, in view of the $g_1^2$ factor
which normalizes the strength of the spectrum, and of eq.
(4.3)).

It becomes now an interesting question to ask whether, in
spite of the above condition, the amplified vacuum
fluctuations can be large enough to seed the dynamo
mechanism which is widely believed to be responsible for the
observed galactic (and extragalactic) magnetic fields [38]. Such
a mechanism would require a primordial magnetic field
coherent over the intergalactic Mpc scale, and with a
minimal strength such that [35]
$$
r(\om_G) \gaq 10^{-34} , \,\,\,\,\,\,\,\,\,\,\,\, \om_G =10^{-14}
Hz \eqno(6.11)
$$
This means, in terms of our parameters,
$$
|y+{x\over2}|>{3\over2}x-{(17+\log_{10}g_1)x \over 25+
{1\over2}\log_{10}g_1} , \,\,\,\,\, x>25+{1\over2}\log_{10}g_1
$$
$$
x(1-\sqrt 3)+|2y+x|>23-0.87 \log_{10}g_1 , \,\,\,\,\,
x<25+{1\over2}\log_{10}g_1 \eqno(6.12)
$$
Surprisingly enough the answer to the previous question is
positive, and this marks an important point in favor of the
string cosmology scenario considered here, as it is in general
quite difficult - if not impossible -  to satisfy the condition
(6.11) in other, more conventional, inflationary scenarios.

The allowed region of parameter space, compatible with the
production of seed fields [eq. (6.12)] in a nearly homogeneous
background [eq. (6.10)], is shown in {\bf Fig. 2} (again for the
reference value $g_1=1$). In the region within the full bold lines
the seed fields are due to modes crossing the horizon in the
dilaton phase, in the region within the thin lines to modes
crossing the horizon in the string phase. In both cases the
background satisfies $y<-x/2$, i.e. $\b>1/2$, so that the whole
allowed region refers to perturbations which are always
growing outside the horizon, even in the string phase.

We may see from {\bf Fig. 2} that the production of seed fields
require a very small value of the dilaton coupling at the
beginning of the string phase,
$$
g_s= e^{\phi_s/2} \laq 10^{-20}\eqno(6.13)
$$
This initial condition is particularly interesting, as it could have
an important impact on the problem of freezing out the
classical oscillations of the dilaton background (work is in
progress). It also requires a long enough duration of the string
phase,
$$
z_s=|\es/\e_1|\gaq 10^{10} \eqno(6.14)
$$
which is not unreasonable, however, when $z_s$ is translated
in cosmic time and string units, $z_s=\exp(\Da t/\la_s)$, namely
$\Da t \gaq 23\la_s$.

\vfill\eject

Also plotted in {\bf Fig. 2}, for comparison, are the allowed
regions for the production of gravitons falling within the
interferometric sensitivity range, taken from the previous
picture. Since there is no overlapping, a signal detected (for
instance) by LIGO would seem to exclude the possibility of
producing seed fields, and vice-versa. Such a conclusion should
not be taken too seriously, however, because the allowed
regions of {\bf Fig. 2} actually define a ``minimal" allowed area,
obtained within the restricted range of parameters compatible
with a linearized description of perturbations. If we drop the
linear approximation, then the allowed area extends to the
``south-western" part of the plane ($x,y$), and an overlap
between electromagnetic and gravitational regions becomes
possible. In that case, however, the perturbative approach
around a homogeneous background could not be valid any
longer, and we would not be able to provide a correct
computation of the spectrum.
\bigskip
\noi
{\bf 6.2. The CMB radiation and its anisotropy}
\bigskip
\noi
In cosmological models based on the low energy string
effective action, the spectrum of scalar and tensor metric
perturbations grows in general too fast with frequency
[3,9,16] to be able to explain the large scale anisotropy
detected by COBE [39,40]. If we insist, however, in looking for
an explanation of the anisotropy in terms of the quantum
fluctuations of some primordial field (amplified by the
background evolution), a possible - even if unconventional -
explanation in a string cosmology context is provided by the
vacuum fluctuations of the electromagnetic field [6].

Consider in fact electromagnetic perturbations, reentering
the horizon ($|k\e|\sim 1\sim \om/H$) after amplification. At
the time of reentry $H^{-1}$ they provide a field coherent over
the horizon scale, which can seed the cosmic magnetic fields, as
discussed in the previous Section. If reentry occurs before the
decoupling era, however, soon after reentry perturbations are
expected to thermalize and homogenize rapidly, because of
their interactions with matter sources in thermal equilibrium.
Modes crossing the horizon after decoupling, on the
contrary,  generate a stochastic perturbation background
whose  spectrum remains frozen until the present time $t_0$,
and that may contribute to the observed inhomogeneities of
the CMB radiation.
In particular, for a complete electromagnetic
origin of the observed
anisotropy, $\Da T/T$, at the present horizon scale,
$\om_0\sim 10^{-18}$Hz, the perturbation amplitude should
satisfy the condition
$$
r(\om_0)\Om_\ga (t_0) \sim (\Da T/T)^2_0
\eqno(6.15)
$$
namely
$$
r(\om_0)\simeq 10^{-6} , \,\,\,\,\,\,\,\,\,\, \om_0= 10^{-18} Hz
\eqno(6.16)
$$

According to our electromagnetic perturbation spectrum [eqs.
(6.8), (6.9)] this condition can be satisfied consistently with the
homogeneity bound (6.10), and without fine-tuning of
parameters, provided the string phase is so long that all scales
inside our present horizon crossed the horizon (for the first
time) during the string phase, i. e. for $\om_0>\om_s$ (or
$z_s>10^{29}$). If we accept this electromagnetic explanation
of the anisotropy, we have then two important consequences.

The first follows from the fact that the peak value of the
spectrum (6.8) is fixed, so that the spectral index $n$, defined
by
$$
r(\om)= g_1^2 \left(\om\over \om_1\right)^{n-1} \eqno (6.17)
$$
can be completely determined as a function of the amplitude
at a given scale. For the horizon scale, in particular, we have
from eqs. (6.15) and (5.3)
$$
n\simeq {25+{5\over 2}\log_{10}g_1- 2 \log_{10}(\Da
T/T)_{\om_0} \over 29 +{1\over 2} \log_{10}g_1} \eqno (6.18)
$$
I have taken explicitly into account here the dependence of
the spectrum on the the present value of the string coupling
$g_1$ (which is illustrated in {\bf Fig. 3}), to stress that such
dependence is very weak, and that our estimate for $n$ from
$\Da T/T$ is quite stable, in spite of the rather large theoretical
uncertainty about $g_1^2$ (nearly two order of magnitude,
recall eq. (4.3)).

In order to match the observed anisotropy, $\Da T/T \sim
10^{-5}$, we obtain from eq. (6.18) (see also {\bf Fig. 3}, where the
relation (6.18) is plotted for three different values of $g_1$)
$$
n \simeq 1.11 - 1.17 \eqno(6.19)
$$
This slightly growing (also called ``blue" spectrum)
is flat enough to be well compatible with the present analyses
of the COBE data [39,40].

\vfill\eject

The second consequence follows from the fact that fixing a
value of $n$ in eq. (6.17) amounts to fix a relation between the
parameters $x$ and $y$ of our background, according to eq.
(6.8). If we accept, in particular, a value of $n$ in the range of
eq. (6.19), then we are in a region of parameter space which is
also compatible with the production of seed fields, according
to eq. (6.11). This means that we are allowed to formulate
cosmological models in which cosmic magnetic fields and CMB
anisotropy have the same common origin, in such a way to
explain (for instance) why the energy density $\r_B$ of the
observed cosmic magnetic fields is of the same order as that
of the CMB radiation: in fact
$$
\r_B \sim \r_\ga \int^{\om_1} r(\om)d(\ln \om) \sim \r_{CMB}
\eqno(6.20)
$$
A coincidence which is otherwise mysterious, to the best of my
knowledge.

It is important to stress that the values of the parameters
leading to eq.(6.19) are also automatically consistent with the
bound following from the presence of strong magnetic fields at
nucleosynthesis time [41], which imposes $r(\om_N)\laq 0.05$
at the scale corresponding to the end of nucleosynthesis,
$\om_N \simeq 10^{-12}$Hz. By comparing photon and graviton
production [eqs. (6.8) and (5.8)] we find, moreover, that for a
background in which $n$ lies in the range (6.19) the graviton
spectrum grows fast enough with frequency ($\Om \sim
\om^{m}$, $m=n+1= 2.11 - 2.17$) to be well compatible with
the pulsar bound (5.12). Note that, with such a value of $m$,
the metric perturbation contribution to the COBE anisotropy is
completely negligible.

It should be mentioned the our electromagnetic perturbation
spectrum, (6.8), (6.9), though obtained from the tree-level in
$g$, lowest order in $\ap$, string effective action (6.1), is
certainly stable with respect to loop corrections when applied
in a range of parameters in which $g=\exp(\phi/2)<<1$, i. e.
the dilaton is deeply inside
the perturbative regime. As to $\ap$ corrections, they could
modify (in principle) the string branch of the spectrum.
However, since we are expanding around the vacuum
background ($F_{\mu\nu}=0$), no higher curvature term which
can be written in the form of powers of the Maxwell
Lagrangian, $(\ap F_{\mu\nu}F^{\mu\nu})^p, p\geq 2$, will
affect the perturbation equations, as long as we are limited to
the linear approximation.

We note, finally, that the class of backgrounds able to provide
an electromagnetic explanation of the CMB anisotropy, can
also account for the production of the CMB radiation itself,
directly from the amplification of the vacuum fluctuations of
the electromagnetic (and other gauge) fields.

Without introducing ``ad hoc" some radiation source,
suppose in fact that the gravi-dilaton background accelerates
up to some maximum (nearly Planckian) scale $H_1$,
corresponding to the peak of the effective potential $V(\e)$ in
the perturbation equations, and then decelerates, with
corresponding decreasing of the potential barrier. This is
enough for the production of a mixture of ultra-relativistic
particles, with a spectrum which is thermal [25] (at a
temperature $T_1\sim H_1a_1/a$) at high frequency
($\om>T_1$), and possibly distorted by parametric
amplification effects at low frequency.

The low frequency part
of the spectrum remains frozen for those particles (like
gravitons and dilatons) which interact only gravitationally, and
then decouple soon after the transition;  it is expected instead
to approach rapidly a thermal distribution for all the other
produced particles which go on interacting among themselves
(and with the background sources) for a long enough time
after the transition. For such particles the total energy, in
critical units, is
$$
\Om_T(t)\sim \left(H_1\over M_p\right)^2
\left(H_1\over H\right)^2\left(a_1\over a\right)^4
\eqno(6.21)
$$
Even if, initially, $\Om_T <1$ (as $H_1<M_p$), such a produced
thermal radiation may thus become dominant ($\Om_T =1$ ),
and identified with the presently observed radiation
background, provided the scale factor, at the beginning of
the decelerated epoch, grows in time more slowly than
$H^{-1/2}$. This is indeed the case, for instance, of the
time-reversed dilaton-dominated solution (4.1), which
expands like $a(t)\sim t^{1/3}$ for $t\ra +\infty$.

It should be stressed, however, that the identification of the
radiation obtained from vacuum fluctuations
with the observed radiation background implies a unique
normalization of all perturbation spectra, $\Om(\om)$, at the
maximum scale $\om_1\sim T_1$: at present time it
imposes $\Om(10^{11}Hz)\sim \Om_T(t_0) \sim 10^{-4}$. On the
other hand we know, from the existing phenomenological
bounds mentioned in this paper, that at low frequencies $\Om<<
10^{-4}$. As a consequence, such a common origin of the CMB
radiation and of its anisotropies can be consistently
implemented only in a background which amplifies fluctuations
with  fast enough growing spectra (which is indeed the case of
the string cosmology scenario discussed here).

\vskip 2 cm
\noi
{\bf 7. Conclusion.}
\bigskip
\noi
In inflationary string cosmology backgrounds perturbations
can be amplified more efficiently than in conventional
inflationary backgrounds, as the perturbation amplitude my
even grow, instead of being constant, outside the horizon. In
some case, like scalar metric perturbations in a dilaton-driven
background, the effects of the growing mode can be gauged
away. But in other cases the growth is physical, and can
prevent a linearized description of perturbations.

In any case, such enhanced amplification is interesting and
worth of further study, as it may lead to phenomenological
consequences which are unexpected in the context of the
standard inflationary scenario. For instance, the production of a
relic graviton background strong enough to be detected by the
large interferometric detectors, or the
production of primordial magnetic fields strong enough to seed
the galactic dynamo. Finally, the possible existence of a
relic stochastic electromagnetic background, due to
the amplification of the vacuum fluctuations of the
electromagnetic field, strong enough to be entirely
responsible for the observed large scale CMB anisotropy.

\vskip 2cm
\noi
{\bf Acknowledgments.}
\bigskip
\noi
I would like to thank R. Brustein, M. Giovannini, V. Mukhanov and G.
Veneziano for a very fruitful and enjoyable collaboration which led to
most of the result reported here.
\vfill\eject

\centerline{\bf References}
\bigskip

\item{1.}L. P. Grishchuk, Sov. Phys. JEPT 40, 409 (1975);

A. A. Starobinski, JEPT Lett. 30, 682 (1979).

\item{2.} M. Gasperini and G.
Veneziano, Mod. Phys. Lett. A8, 3701 (1993)

\item{3.}R. Brustein, M. Gasperini, M. Giovannini, V. F. Mukhanov
and G.
Veneziano, {\it Metric perturbations in dilaton-driven inflation},
Phys. Rev. D (1995), in press (hep-th/9501066).

\item{4.}R. Brustein, M. Gasperini, M. Giovannini and G.
Veneziano, {\it Relic gravity wave from string cosmology},
CERN-TH./95-144 (June 1995)

\item{5.}M. Gasperini, M. Giovannini and G. Veneziano, {\it
Primordial magnetic fields from string cosmology},
CERN-TH/95-85 (hep-th/9504083)

\item{6.}M. Gasperini, M. Giovannini and G. Veneziano, {\it
Electromagnetic origin of the CMB anisotropy in string
cosmology},
CERN-TH/95-102 (astro-ph/9505041)

\item{7.}G. Veneziano, Phys. Lett. B265 , 287 (1991)

\item{8.}M. Gasperini
and G. Veneziano, Astropart. Phys. 1, 317 (1993)

\item{9.}M. Gasperini and G. Veneziano, Phys. Rev. D50,
2519 (1994)

\item{10.}R. Brustein and G. Veneziano, Phys. Lett. B329,
429 (1994);

N. Kaloper, R. Madden and K. A. Olive, {\it Towards a singularity-free
inflation-

ary universe?}, UMN-TH-1333/95 (June 1995)

\item{11.}M. Gasperini, in  ``Proc. of the 2nd Journ\'ee
Cosmologie" (Observatoire de Paris, June 1994), ed. by N.
Sanchez and H. de Vega  (World Scientific, Singapore), p.429

\item{12.}D. Shadev, Phys. Lett. B317, 155 (1984);

R. B. Abbott, S. M. Barr and S. D. Ellis, Phys. Rev. D30, 720 (1984);

E. W. Kolb, D. Lindley and D. Seckel, Phys. Rev. D30, 1205 (1984);

F. Lucchin and S. Matarrese, Phys. Lett. B164, 282 (1985).

\item{13.}A. A. Starobinski, Rel. Astr. Cosm., Byel. SSR Ac. Sci. Minsk
(1976), p.55 (in russian)

\item{14} R. B. Abbott, B. Bednarz and S. D. Ellis, Phys. Rev.
D33, 2147 (1986)

\item{15.}V. Mukhanov, H. A. Feldman and R. Brandenberger,
Phys. Rep. 215, 203 (1992)

 \item{16.}M. Gasperini and M. Giovannini,
Phys. Rev. D47, 1529 (1992)

\item{17.}L. P. Grishchuk and Y. V. Sidorov,
Phys. Rev. D42, 3413 (1990)

\item{18.}J. D. Barrow, J. P. Mimoso and M. R. de Garcia Maia,
Phys. Rev. D48, 3630 (1993)

\item{19.}M. Gasperini, M. Giovannini, K. A. Meissner and G.
Veneziano; {\it Evolution of a string network in backgrounds
with rolling horizons} (CERN-TH/95-40), to appear in
``New developments in string
gravity and physics at the Planck energy scale", ed. by N. Sanchez
(World Scientific, Singapore, 1995)

\item{20.}S. W. Hawking, Astrophys. J. 145, 544 (1966)

\item{21.}A. R. Liddle and D. H. Lyth, Phys. Rep. 231, 1 (1993)

\item{22.}M. Bruni, G. F. R.
Ellis and P. K. S. Dunsby, Class. Quantum
Grav. 9, 921 (1992)

\item{23.}M. Gasperini, Phys. Lett. B327, 214 (1994)

\item{24.}K. A. Meissner and G. Veneziano, Phys. Lett. B267,
 33 (1991); Mod. Phys. Lett. A6, 3397 (1991); M. Gasperini and G.
Veneziano, Phys. Lett. B277, 256 (1992).

\item{25.}L. Parker, Nature 261, 20 (1976)

\item{26.}R. Brustein, M. Gasperini and M. Giovannini, {\it
Possible common origin of primordial perturbations and of the
cosmic microwave background}, Essay written for the 1995
Awards for Essays on Gravitation (Gravity Research Foundation,
Wellesley Hills, Ma), and selected for Honorable Mention
(unpublished)

\item{27.} A. Abramovici et al., Science 256, 325 (1992)

\item{28.} B. Caron et. al.,  {\it Status of the VIRGO
experiment}, Lapp-Exp-94-15

\item{29.}G. Veneziano, in ``Proc. of the PASCOS '94
Conference"   (Syracuse, N.Y., 1994) (CERN-TH.7502/94)

 \item{30.} K. S. Thorne, in ``300 Years of Gravitation", ed. by S.
W. Hawking and W. Israel (Cambridge Univ. Press, Cambridge,
1987)

\item{31.}D. R. Stinebring et al., Phys. Rev. Lett. 65, 285 (1990)

\item{32.}V. F. Schwarztmann, JEPT Letters 9, 184 (1969)

\item{33.}P. Astone, J. A. Lobo and B. F. Schutz, Class.
Quantum Grav. 1, 2093 (1994)

\item{34.} F. Pegoraro, E. Picasso, L. Radicati, J. Phys.
A1, 1949 (1978);  C. M. Caves, Phys. Lett. B80, 323 (1979);
C. E. Reece et al., Phys. Lett A104, 341 (1984).

\item{35.} M. S. Turner and L. M. Widrow,
Phys. Rev. D37, 2743 (1988).

\item{36.} B. Ratra, Astrophys. J. Lett. 391, L1 (1992);
A. D. Dolgov, Phys. Rev. D48, 2499, (1993).

\item{37.}D. Lemoine and M. Lemoine, {\it Primordial magnetic
fields  in string cosmology}, April 1995

\item{38.} E. N. Parker,
``Cosmical Magnetic fields" (Clarendon, Oxford,
England, 1979)

\item{39.}G. F. Smoot et al., Astrophys. J. 396, L1 (1992)

\item{40.}C. L. Bennett et al., Astrophys. J. 430, 423 (1994)

\item{41.}D. Grasso and H. R. Rubinstein, Astropart. Phys. 3, 95
(1995)

\end